\documentclass[mnsc]{amsart}
\usepackage{graphics,amsmath,amssymb,epsfig,stmaryrd,color,amsfonts}
\usepackage{hyperref}

\hypersetup{
    colorlinks=true,
    urlcolor=blue,
    linkcolor=blue,
    citecolor=blue
}

\newtheorem{theorem}{Theorem}
\theoremstyle{plain}

\newtheorem{assumption}{Assumption}

\newtheorem{corollary}{Corollary}

\newtheorem{definition}{Definition}

\newtheorem{proposition}{Proposition}
\newtheorem{remark}{Remark}

\numberwithin{equation}{section}

\begin{document}

\title{Local utility and multivariate risk aversion}

\author[Charpentier]{Arthur Charpentier$^1$}
\author[Galichon]{Alfred Galichon$^2$}
\author[Henry]{Marc Henry$^3$}

\thanks{This version: January 17, 2015. The authors thank the editor and two anonymous referee for numerous very helpful comments and suggestions and participants at RUD 2012 for helpful comments. Charpentier received financial support from NSERC and the Research Chairs AXA/FdR. Galichon received financial support from the Research Chairs AXA ``Assurance et Risques Majeurs'', EDF-Calyon ``Finance and D\'eveloppement Durable'' and from FiME, Laboratoire de Finance des March\'es de l'Energie (www.fime-lab.org). Parts of this paper were written while
Henry was visiting the University of Tokyo Graduate School of
Economics and
he gratefully acknowledges his hosts and the
CIRJE for their support. 
Henry also received financial support from SSHRC Grants 410-2010-242 and 435-2013-0292 and NSERC Grant
356491-2013.\\
$^1$Arthur Charpentier, UQAM \& Univerist\'e de Rennes 1, charpentier.arthur@uqam.ca.\\
$^2$Alfred Galichon, Sciences-Po, Paris, alfred.galichon@sciences-po.fr.\\
$^3$Marc Henry, The Pennsylvania State University, marc.henry@psu.edu} 

\begin{abstract}
We revisit Machina's local utility as a tool to analyze attitudes to multivariate risks. We show that for non-expected utility maximizers choosing between multivariate prospects, aversion to multivariate mean preserving increases in risk is equivalent to the concavity of the local utility functions, thereby generalizing Machina's result in \cite{Machina:1982}. To analyze comparative risk attitudes within the multivariate extension of rank dependent expected utility of Galichon and Henry \cite{GH:2011}, we extend Quiggin's monotone mean and utility preserving increases in risk and show that the useful characterization given in Landsberger and Meilijson \cite{LM:94} still holds in the multivariate case.\\


\noindent{\textbf{MSC class}: 00A06}\\

\end{abstract}%

\maketitle

\newpage
\section*{Introduction}
One of the many appealing features of expected utility theory is the characterization of attitudes towards risk through the shape of the utility function. Following extensive evidence of violations of the independence axiom which delivers linearity in probabilities of the functional characterizing preferences over risky prospects, most notably the celebrated Allais paradox \cite{Allais:53}, Machina showed in \cite{Machina:1982}, \cite{Machina:1982b} that smoothness of the
preference functional was sufficient to recover representability of risk attitudes through a local approximation, which he called {\em local utility function}.
Parallel to the study of risk attitudes in generalized expected utility theories, Stiglitz \cite{Stiglitz:69} and Kihlstrom and Mirman \cite{KM:74} analyzed attitudes to the combination
of income risk and price risk in preferences over multiple commodities within the expected utility framework. This paper is concerned with non expected utility analysis of attitudes to multivariate risks. So far, three approaches have emerged to analyze attitudes to multivariate risks without the independence axiom in \cite{Yaari:86}, \cite{SS:93} and \cite{GKP:92}. All three apply dimension reduction devices to preferences over multivariate prospects. Yaari \cite{Yaari:86}
considers rank dependent utility over multivariate prospects with stochastically independent components only; Safra and Segal \cite{SS:93} show additive separability of the local utility function under a property they call {\em dominance} (equivalent to the notion of {\em correlation neutrality} in \cite{ET:80}) and Grant, Kajii and Polak \cite{GKP:92} show that under a property they call {\em degenerate independence}, preferences over uncertain multivariate prospects can be fully recovered from preferences over uncertain income and preferences over deterministic multivariate outcomes.
We consider the general case, where attitudes to income risk and price risk cannot be separated in this way and show that in general smooth preferences over multivariate prospects, the main result of Machina \cite{Machina:1982} still holds, and aversion to increases in risk is equivalent to concavity of the local utility function. The proof relies on the martingale characterization of increasing risk in Galichon and Henry \cite{GH:2011}. A special case of this result appears in Galichon and Henry \cite{GH:2011}, who derive the family of local utility functions in a multivariate rank dependent utility model under aversion to multivariate mean preserving increases in risk. Machina also showed in \cite{Machina:1982} that interpersonal comparisons of risk aversion can be characterized by properties of the local utility function. Karni generalizes in \cite{Karni:89} the equivalence between decreasing certainty equivalents and concave transformations of the local utility functions to smooth preferences over multivariate prospects. To complement this result, we extend the notion of compensated spread to multivariate prospects and generalize the characterization of Quiggin's monotone increases in risk \cite{Quiggin:92} as mean preserving comonotonic spreads in \cite{LM:94}. This allows us to recover a multivariate version of Landsberger and Meilijson's seminal result on the efficiency of partial insurance contracts for monotone mean preserving reductions in risk in \cite{LM:94}. We also generalize Quiggin's notion of pessimism and characterize pessimistic decision functionals by the shape of their local utility function. We apply these notions to interpersonal comparison of risk aversion within the multivariate rank dependent model of \cite{GH:2011} and we show that pessimism is equivalent to weak risk aversion in that framework.

The rest of the paper is organized as follows. Section~\ref{section: local utility} defines local utility. Section~\ref{section: risk aversion} shows that aversion to mean preserving increases in risk is equivalent to concavity of the local utility functions and Section~\ref{section: comparisons} extends Quiggin's monotone mean preserving increases in risk and Section~\ref{subsection: IRA} applies it to interpersonal comparisons of risk aversion within the multivariate rank dependent utility model. The last section concludes.

\subsection*{Notation and basic definitions}

Let $({S},{\mathcal{F}},\mathbb{P})$ be a non-atomic probability space. Let $%
X:{S}\rightarrow {\mathbb{R}}^{d}$ be a random vector. We denote the
cumulative distribution function of $X$ by $F_X$.
$\mathbb{E}$ is the expectation operator with respect to $\mathbb{P}$. For $%
x $ and $y$ in ${\mathbb{R}}^{d}$, let $x\cdot y$ be the standard
scalar product of $x$ and $y$, and $\left\Vert x\right\Vert ^{2}$
the Euclidian norm of $x$. We denote by $X=_d\mu$ the
fact that the distribution of $X$ is $\mu$ and by
$X=_{d}Y$ the fact that $X$ and $Y$ have
the same distribution. $Q_{X}$ denotes the
quantile function of distribution $X$.
In dimension 1, this is defined for all $t\in \lbrack
0,1] $ by $Q_{X}(t)=\inf_{x\in \mathbb{R}}\{\mathrm{Pr}(X\leq x)>t\}$. In
larger dimensions, it is defined in Definition~\ref{definition: generalized
quantiles} of Section~\ref{subsection: multivariate quantiles and comonotonicity} below. We
call $L_d^1$ the set of integrable random vectors of dimension $d$ and $L_{d}^{2}$ the set of random vectors $X$ of dimension $d$ such that $%
\mathbb{E}\left\Vert X\right\Vert ^{2}<\infty $. We
denote by $\mathcal{D}$ the subset of $L_{d}^{2}$ containing random
vectors with a density relative to Lebesgue measure.
A functional $\Phi$ on $L_d^2$ is
called upper semi-continuous (denoted u.s.c.) if  for any real
number $\alpha$, $\{X\in L_d^2:\;\Phi(X)<\alpha\}$ is open. A
functional $\Phi$ is lower semi-continuous (l.s.c.) if $-\Phi$ is
upper semi-continuous. $\Phi$ is called law-invariant if $\Phi(X)=\Phi(\tilde X)$ whenever $\tilde X=_dX$. By a slight abuse of notation, when $\Phi$ is law invariant, $\Phi(F_X)$ will be used to denote $\Phi(X)$. For a convex lower semi-continuous function $V:\mathbb{R}^{d}\rightarrow \mathbb{%
R}$, we denote by $\nabla V$ its gradient (equal to the vector of
partial derivatives).

\section{Local Utility}
\label{section: local utility}
We consider decision makers choosing among distributions functions on $\mathbb R^d$ with finite mean. We assume that the decision makers' preferences are given as a complete, reflexive and transitive binary relation represented by a real valued functional $\Phi$, which is continuous relative to the topology of convergence in distribution. 
Suppose further that $\Phi$ is G\^ateaux differentiable.
\begin{assumption}[Local Utility] The following properties hold.
\begin{enumerate}
\item $\Phi$ is continuous with respect the topology of weak convergence of probability measures.
\item For each distribution function $F$ on $\mathbb R^d$, there is a function $x\mapsto U_\Phi (x,F)$ such that, for each distribution function $G$, 
\begin{eqnarray*}
\frac{d}{dt}\Phi\left[ (1-t)F+tG\right]\bigr\rvert_{0^+}=\int U_\Phi (x,F)\left[dG(x)-dF(x)\right].
\end{eqnarray*}
\end{enumerate}
The function $U_\Phi (x;F)$ thus defined is called local utility function relative to $\Phi$ at $F$. \label{ass:gat}
\end{assumption}
Since expected utility preferences are linear in probabilities, the local utility of an expected utility decision maker is constant and equal to her utility function.
Theorem~1 in \cite{Machina:1982} and its extension to G\^ateaux differentiability in \cite{CKS:87} for the special case of Rank Dependent Utility, show that smooth preference functionals are monotonic if and only if their local utility functions are increasing. This can be extended to the case of multivariate prospects.
\begin{definition}[Stochastic dominance] A distribution $F$ is said to dominate
stochastically a distribution $G$ (denoted $F\succsim_{SD}G$) if there exist $\tilde X=_dF$ and $\tilde Y=_dG$
such that $\tilde X\geq\tilde Y$ almost surely, where $\geq$ denotes componentwise order in $\mathbb R^d$.\end{definition} A preference functional is said to preserve stochastic dominance if stochastically dominant prospects are always preferred. If the preference functional $\Phi$ is law invariant and monotonic, in the sense that $\Phi(X)\geq\Phi(Y)$ when $X$ yields larger outcomes than $Y$ in almost all states, then it preserves stochastic dominance. The proof of Theorem~1 of \cite{Machina:1982} is dimension free and therefore, a Fr\'echet differentiable preference functional preserves stochastic dominance if and only if the utility function is non decreasing. The proposed extension to G\^ateaux differentiable functionals in \cite{CKS:87} is specific to Rank Dependent Utility, however.

If in addition, the decision maker is indifferent to correlation increasing transfers, or correlation neutral according to the terminology of \cite{ET:80},
then Safra and Segal show in \cite{SS:93} that the local utility functions are additively separable, namely that $U_\Phi (x;F)=\sum_{j=1}^dU_j(x_j;F)$, where $x_j$ is the $j$-th component of the outcome $x\in\mathbb R^d$. Yaari's rank dependent utility maximizers over stochastically independent $d$-dimensional risks in \cite{Yaari:86} are represented by \begin{equation}\label{equation: Yaari}\Phi(X)=\sum_{i=1}^d\alpha_i\int_0^1\phi_i(u)Q_{X_i}(t)dt,\end{equation} where $Q_{X_i}$ is the quantile function of component $X_i$ of the risk $X$, the $\phi_i$'s, $i=1,\ldots,d$, are non-negative functions on $[0,1]$ (quantile weights interpreted as probability distortions) and the $\alpha_i$'s, $i=1,\ldots,d$, are positive weights. The local utility of decision maker $\Phi$ is given by \begin{equation}\label{equation: Yaari LU} U_\Phi (x;F)=\sum_{i=1}^d\alpha_i\int^{x_i}\phi_i(F_i(z))dz,\end{equation} where $F_i$ is the $i$-th marginal of distribution $F$ (see for instance Section~4 of \cite{Segal:87}).

\section{Local utility and mean preserving increases in risk}
\label{section: risk aversion}
We now show that attitude to risk with smooth preference over multivariate prospects
can be characterized by the shape of local utilities, as was proved in the case of univariate risks in Theorem~2 of \cite{Machina:1982}. The latter shows
that aversion to mean preserving increases in risk is equivalent to concavity of local utility functions. Extending this result to preferences over multivariate prospects calls for a generalization of the notion of mean preserving increase in risk
proposed in \cite{RS:70}.
\begin{definition}[Mean preserving increase in risk]\label{definition: MPIR}
A distribution $G$ is called a mean preserving increase in risk (hereafter MPIR) of a distribution $F$, denoted $G\succsim_{MPIR}F$, if either of the following equivalent statements hold.
\begin{itemize}\item[(a)] For all concave functions $f$ on $\mathbb R^d$, $\mathbb \int fdF\geq\mathbb \int fdG$.\item[(b)] There exists $Y=_d G$  and $X=_d F$ such that $(X,Y)$ is a martingale, i.e., $\mathbb E[Y\vert X]=X$.
\end{itemize}
\end{definition}
The equivalence between (a) and (b), Theorem 7.A.1 in \cite{SS:2007}, is due to \cite{Strassen:65} and the interpretation as an increase in risk is the same as in \cite{RS:70} for the univariate case. When the domain is restricted to $\mathcal D$, \cite{GH:2011} show that (a) and (b) are also equivalent to (c): For all u.s.c. law invariant concave functionals $\Psi$ on $\mathcal D$ and any $X=_d F$ and $Y=_d G$, $\Psi(X)\geq\Psi(Y)$. An immediate corollary of the latter is that cardinal risk aversion, i.e., concavity of the functional representing preferences, implies ordinal risk aversion, in the sense of aversion to mean preserving increases in risk.
We can now state the main result of this section, which is a direct generalization of Theorems~2 and~3 of \cite{Machina:1982}.
\begin{theorem}[Risk aversion and local utility]\label{theorem: Schur}
Let $\Phi$ be a preference functional satisfying Assumption~\ref{ass:gat}. Then the following statements are equivalent. (i) $\Phi$ is risk averse, i.e., $\Phi(F)\geq\Phi(G)$ when $G$ is an MPIR of $F$, (ii) $U_\Phi (\cdot;F)$ is a concave function for all $F$ and (iii) For arbitrary prospects $F$ and $F^\ast$ and any $\alpha\in[0,1]$, $\Phi(\alpha F+(1-\alpha)G_{\mu_{F^\ast}})\geq\Phi(\alpha F+(1-\alpha)F^\ast)$, where $\mu_F$ is the mean of $F$ and $G_\mu$ is the degenerate distribution at $\mu$.
\end{theorem}

\begin{proof}[Proof of Theorem~\ref{theorem: Schur}]
(i) $\Leftrightarrow$ (iii): Using the martingale difference characterization MPIR, it is easy to show that $\alpha F+(1-\alpha)F^\ast$ is a mean preserving increase in risk relative to  $\alpha F+(1-\alpha)G_{\mu_{F^\ast}}$, so that monotonicity with respect to MPIR implies (iii). 

The converse is proved in the following way. 
A probability measure $Q$ on $\mathbb R^d$ is called an elementary fusion of a probability measure $P$ (in the terminology of \cite{EH:92}) if there is a set $A$ and $\beta\in[0,1]$ such that
\[Q=P_{|A^c}+\beta P_{|A}+(1-\beta)P(A)\delta_{\mu_A},\]
where $\mu_A$ is the mean of $P_{|A}$. A probability measure $P_n$ is called a simple fusion of a probability measure $P$ if $P_n$ can be obtained from $P$ as the result of a sequence of $n$ elementary fusions.

We first show that under Condition (iii), $\Phi(Q)\geq\Phi(P)$ whenever $Q$ is an elementary fusion of $P$.
Take $P$ a probability measure, $\beta\in[0,1]$ and a set $A$ and find $\alpha,$ $Q$ and $Q^\ast$ with mean $\mu^\ast$ such that 
\begin{eqnarray*}
&&P=\alpha Q+(1-\alpha)Q^*\;\mbox{ and }\\
&&P_{|A^c}+\beta P_{|A}+(1-\beta)P(A)\delta_{\mu_A}=\alpha Q+(1-\alpha)\delta_{\mu^*}.
\end{eqnarray*}
This implies
$\alpha Q=P_{|A^c}+\beta P_{|A}$ and 
$(1-\alpha)\delta_{\mu^*}=(1-\beta)P(A)\delta_{\mu_A}.$
Hence,
\begin{eqnarray*}
\alpha&=&1-(1-\beta)P(A),\\
Q&=&\frac{1}{1-(1-\beta)P(A)}\left[ P_{|A^c}+\beta P_{|A} \right],\\
Q^*&=&\frac{1}{(1-\beta)P(A)}\left[ P- P_{|A^c}-\beta P_{|A} \right].
\end{eqnarray*}
There remains to check that $Q^\ast$ is centered at $\mu_A$.
Indeed, the mean of $Q^*$ is \[\frac{1}{(1-\beta)P(A)}\left[ P(A)\mu_A+P(A^c)\mu_{A^c}- P(A^c)\mu_{A^c}-\beta P(A)\mu_A \right]=\mu_A.\] 
By (iii), we know that $\Phi(\alpha Q+(1-\alpha)Q^*)\leq\Phi(\alpha Q+(1-\alpha)\delta_{\mu^*})$ hence we have $\Phi(P)\leq\Phi(Q)$ for any $Q$ elementary fusion of any $P$ with finite mean. 

By the continuity of $\Phi$ from Assumption~\ref{ass:gat} (1), this implies that $\Phi(Q)\geq\Phi(P)$, whenever $Q$ is the limit of a sequence of simple fusions of $P$. By the equivalence between (i) and (ii) in Theorem~4.1, page 47 of \cite{EH:92}, $\Phi(Q)\geq\Phi(P)$ whenever $P$ is an MPIR of $Q$.
The implication (iii) $\Longrightarrow$ (i) follows.  

(ii) $\Longrightarrow$ (iii):
Write $F_Y=F$ and $F_X=F^\ast$. Consider the following two lotteries, so that $Z=_d \alpha F+(1-\alpha)F^*$ and $\tilde Z=_d \alpha F+(1-\alpha)G_{\mu_{F^*}} $,
\begin{equation}\label{Eq:lottery}
Z\begin{matrix}
{_\alpha} \\
\nearrow \\
\searrow  \\
^{1-\alpha}
\end{matrix}
\begin{matrix}
F \\
  \\
F^*
\end{matrix}
\quad \text{ and }\quad
\tilde X\begin{matrix}
{_\alpha} \\
\nearrow \\
\searrow  \\
^{1-\alpha}
\end{matrix}
\begin{matrix}
F \\
  \\
G_{\mu_{F^*}}
\end{matrix}
\end{equation}
Given $\varepsilon\in[0,1]$, consider $Z_\varepsilon$ a mixture between $Z$ and $\tilde Z$, with weights $\varepsilon$ and $1-\varepsilon$, and let $Z_\varepsilon =_dF_\varepsilon$,
$$
F_\varepsilon = \alpha F+(1-\alpha)
[(1-\varepsilon)G_{\mu_{F^*}}+\varepsilon F^*].
$$
For $h\geq 0$, note that
\begin{equation}\label{equation:Feps}
F_{\varepsilon+h} = F_{\varepsilon} +(1-\alpha)h[F^*-G_{\mu_{F^*}}].
\end{equation}
If we substitute
$$
G_{\mu_{F^*}}=\frac{1}{(1-\alpha)(1-\varepsilon)}[F_\varepsilon-\alpha F-(1-\alpha)\varepsilon F^*]
$$
in equation~(\ref{equation:Feps}), we get
$$
F_{\varepsilon+h} = \left[1-\frac{h}{1-\varepsilon}\right] F_{\varepsilon} + \frac{h}{1-\varepsilon}
\left[
\alpha F +(1-\alpha) F^*
\right],
$$
so that
$$
\Phi(F_{\varepsilon+h})-\Phi(F_{\varepsilon})=
\Phi\left(\left[1-\frac{h}{1-\varepsilon}\right]
F_{\varepsilon} + \frac{h}{1-\varepsilon}
\left[
\alpha F +(1-\alpha) F^*\right]\right)
-\Phi\left(
F_{\varepsilon}
\right).
$$
Let $H=\left[
\alpha F +(1-\alpha) F^*\right]$, so that this expression becomes
$$
\Phi((1-\eta)F_{\varepsilon}+\eta H)-\Phi\left(
F_{\varepsilon}
\right)$$
with $\eta=h/(1-\varepsilon)$.
Now,
\begin{eqnarray*}
\Phi((1-\eta)F_{\varepsilon}+\eta H)-\Phi\left(
F_{\varepsilon}\right) &=& \int U_\Phi (x;F_{\varepsilon}) d[(1-\eta)F_{\varepsilon}+\eta H-F_{\varepsilon}]+o(h)\\
&=&\eta \int U_\Phi (x;F_{\varepsilon}) d[\left[
\alpha F +(1-\alpha) F^*\right]-F_{\varepsilon}]+o(h)
\end{eqnarray*} which equals
\begin{eqnarray*}
\frac{h}{1-\varepsilon}\left[
\int U_\Phi (x;F_\varepsilon) d[\alpha F+(1-\alpha) F^*]
-\int U_\Phi (x;F_\varepsilon)dF_\varepsilon
\right]
+o\left(h\right).
\end{eqnarray*}
Since
$$
\frac{d}{d \varepsilon} \Phi(F_\varepsilon)
=\lim_{h\rightarrow 0}\frac{\Phi(F_{\varepsilon+h})-\Phi(F_{\varepsilon})}{h},
$$
using $F_\varepsilon=\alpha F+(1-\alpha)[\varepsilon F^\ast+(1-\varepsilon)G_{\mu_{F^\ast}}]$, we find
$$
\frac{d}{d \varepsilon} \Phi(F_\varepsilon)
= \lim_{h\rightarrow 0} \frac{1}{h}\left[ h (1-\alpha)
\left(
\int U_\Phi (x;F_\varepsilon)dF^* - U_\Phi (\mu_{F^*};F_\varepsilon)\right)
\right]
$$
i.e.,
$$
\frac{d}{d \varepsilon} \Phi(F_\varepsilon)
= (1-\alpha)\left[
\int U_\Phi (x;F_\varepsilon)dF^* - U_\Phi (\mu_{F^*};F_\varepsilon)
\right]
\leq 0
$$ by Jensen's inequality, 
since $U_\Phi (\cdot,F_\varepsilon)$ is a concave function.
Hence, we obtain that $\Phi(F_0)\geq \Phi(F_1)$, i.e.,
$$
\Phi(\alpha F+(1-\alpha) F^*) \geq \Phi(\alpha F+(1-\alpha) G_{\mu_F^*}).
$$

(iii) $\Longrightarrow$ (ii): 
By (iii), we have $\Phi(\alpha F+(1-\alpha)G_{\mu_{F^\ast}})\geq\Phi(\alpha F+(1-\alpha)F^\ast)$, which yields, by G\^ateaux differentiability, $\int U_\Phi (x,F)dG_{\mu_{F^\ast}}(x)\geq\int U_\Phi (x,F)dF^\ast(x)$. Hence, $U_\Phi (\mu,F)\geq\int U_\Phi (x,F)dF^\ast(x)$, which implies concavity of $U_\Phi (\cdot,F)$, as required.
\end{proof}

Using the local utility, we extend insights from the vast literature on multivariate risk taking (see for instance \cite{ERS:2007} and references therein) to non expected utility preference functions. We can also define a full insurance premium for preferences over multivariate prospects. Let $F$ be a prospect evaluated by a decision maker with smooth preferences as $\Phi(F)$. A full insurance premium can be defined as an element of the set of vectors $\pi\in\mathbb R^d$ satisfying $\Phi(F)=U_\Phi (\mu-\pi;F)$, where $\mu$ is the mean of~$F$.

\section{Multivariate mean preserving increases in risk}
\label{section: comparisons}

In Section~\ref{sub:Quig}, we consider univariate risks ($d=1$) and characterize local utility of decision makers that are averse to Quiggin's monotone mean preserving increases in risk, using a celebrated result of Landsberger and Meilijson \cite{LM:94}. In Section~\ref{subsection:multiMMPIR}, we extend the latter to the multivariate case, order to provide the multivariate equivalent of aversion to monotone mean preserving increases in risk and its local utility characterization.

\subsection{Aversion to monotone mean preserving increases in risk} 
\label{sub:Quig}
In \cite{Quiggin:92}, Quiggin shows that the notion of mean preserving increases in risk is too weak to coherently order rank dependent utility maximizers according to increasing risk aversion. Quiggin \cite{Quiggin:92} shows that the notion of monotone mean preserving increases in risk (Monotone MPIR) is the weakest stochastic ordering that achieves a coherent ranking of risk aversion in the rank dependent utility framework. Monotone MPIR is the mean preserving version of Bickel-Lehmann dispersion (\cite{BL:76},\cite{BL:79}), which we now define. 
\begin{definition}[Bickel-Lehmann Dispersion]
\label{definition: BL dispersion}
Let $Q_X$ and $Q_Y$ be the quantile functions of the random variables $X$ and $Y$. $X$ is said to be {\em Bickel-Lehmann less dispersed}, denoted  $X\precsim_{BL}Y$, if $Q_Y(u)-Q_X(u)$ is a nondecreasing function of $u$ on $(0,1)$. The mean preserving version is called monotone mean preserving increase in risk (hereafter MMPIR) and denoted $\precsim_{MMPIR}$.
\end{definition}
MMPIR is a stronger ordering than MPIR in the sense that $X\precsim_{MMPIR}Y$ implies $X\precsim_{MPIR}Y$ since it is shown in \cite{DS:74} that an MPIR can be obtained as the limit of a sequence of simple mean preserving spreads $Y$ of $X$, defined by $Q_Y(u)-Q_X(u)$ non-positive below some $u_0\in[0,1]$ and non-negative above $u_0$.
\cite{Quiggin:92} relates MMPIR aversion of a rank dependent utility decision maker to a notion he calls {\em pessimism}. Aversion to MMPIR is defined in the usual way as follows.
\begin{definition}[Risk aversion]
A preference functional $\Phi$ over random prospects is called {\em averse to monotone mean preserving increases in risk} if and only if $X\precsim_{MMPIR}Y$ implies $\Phi(X)\geq\Phi(Y)$.
\end{definition}
Consider a decision maker with preference relation characterized by the functional defined for each prospect distribution $F$ by \begin{eqnarray}\label{eq: RDU}\Phi(F)=\int f(1-F(x))dx\end{eqnarray}
with $f(0)=0$, $f(1)=1$ and $f$ non decreasing. Then Theorem~3 of \cite{CCM:2004} shows that aversion to MMPIR is equivalent to $f(u)\leq u$ for each $u\in[0,1]$. Since the local utility associated with $\Phi$ is $x\mapsto U_\Phi (x,F)= \int^xf^\prime(1-F(z))dz$, aversion to MMPIR can be characterized with the local utility. We now generalize this local utility characterization of MMPIR aversion beyond rank dependent utility functionals to all preference functionals
that admit a local utility. For the purpose of this characterization, we strengthen the differentiability requirement of Assumption~\ref{ass:gat} to Fr\'echet differentiability with smooth local utility.
\begin{assumption}[Smooth local utility]\label{ass:fre}
For each distribution function $F$ on $\mathbb R$, there exists a differentiable function $x\mapsto U_\Phi(x,F)$, such that for all distribution $G$,
\[
\Phi(G)-\Phi(F)=\int U_\Phi (x,F)[dG(x)-dF(x)]+o(d(F,G)),
\]
where $d$ is the 1-Wasserstein distance
\[
d(F,G):=\inf\left\{ \mathbb E\vert X-Y\vert;\;X=_dF,\;Y=_d G\right\},
 \] 
 which metrizes the topology of convergence in distribution (see Theorem~6.9 of \cite{Villani:2009}).
\end{assumption}
\begin{theorem}[Local utility of MMPIR averse decision makers]
\label{thm:BL1}
Let $\Phi$ be a preference functional on $L^1$ distributions satisfying Assumption~\ref{ass:fre}. $\Phi$ is MMPIR averse if and only if 
\[
\int U_\Phi^\prime(x,F)\delta(x)dF(x)\leq0,
\] 
for all $F$ and all non decreasing functions $\delta$, such that $\int\delta dF=0$ and $\int\vert\delta\vert dF<\infty$.\label{theorem: MMPIR aversion}
\end{theorem}
\begin{remark} $\mbox{ }$ Note that $\delta$ can be chosen equal to $y\mapsto\delta(y)=1\{y> x\}-[1-F_X(x)]$ for any $x\in\mathbb R$ and in the special case of rank dependent utility functional (\ref{eq: RDU}), the characterization above is equivalent to $f(1-F_X(x))\leq 1-F_X(x)$ for all $x$ and $X$, which is equivalent to $f(u)\leq u$ for all $u\in[0,1]$ as mentioned previously.\end{remark}

In Proposition~2 of \cite{LM:94}, Landsberger and Meilijson give a characterization of Bickel-Lehmann dispersion in the spirit of the characterization of MPIR given in the equivalence between (a) and (b) of Proposition~\ref{definition: MPIR}. In the latter, MPIR increases are characterized by the addition of noise, whereas in the former MMPIR are characterized by the addition of a zero mean comonotonic variable.
\begin{proposition}[Landsberger-Meilijson]\label{proposition: LM}
A random variable $X$ has Bickel-Lehmann less dispersed distribution than a random variable $Y$ if and only iff there exists $Z$ comonotonic with $X$ such that $Y=_dX+Z$.
\end{proposition}
Using Proposition~\ref{proposition: LM}, we can prove Theorem~\ref{theorem: MMPIR aversion}.
\begin{proof}[Proof of Theorem~\ref{theorem: MMPIR aversion}]
From Proposition~\ref{proposition: LM}, $\Phi$ is MMPIR averse if and only if $\Phi(X+Z)-\Phi(X)\leq0$ for any $(X,Z)$ comonotonic and $\mathbb EZ=0$. Take, therefore, $X$ and $Z$ two comonotonic random variables, with $Z$ in $L^1$ and centered. Call $F$ the distribution of $X$ and $F_\varepsilon$ the distribution of $X+\varepsilon Z$, for $\varepsilon>0$. Note that $\varepsilon Z$ and $X$ are also comonotonic. By Assumption~\ref{ass:fre}, since $\mathbb E\vert X+\varepsilon Z-X\vert=O(\varepsilon)$, \begin{eqnarray*}\Phi(F_\varepsilon)-\Phi(F)&=&\int U_\Phi (x,F)[dF_\varepsilon(x)-dF(x)]+o(\varepsilon)\\
&=&\int_0^1 U_\Phi (Q_{X+\varepsilon Z}(u))-U_\Phi (Q_X(u)]du+o(\varepsilon)\\
&=&\int_0^1 U_\Phi (Q_{X}+Q_{\varepsilon Z}(u))-U_\Phi (Q_X(u)]du+o(\varepsilon)\\
&=&\varepsilon\int_0^1U_\Phi^\prime(Q_X(u),F)Q_{Z}(u)du+o(\varepsilon)\end{eqnarray*} where the penultimate equation holds because the quantile function is comonotonic additive and the last equation holds because $Z$ is integrable. Therefore \[\int_0^1 U_\Phi^\prime(Q_X(u),F)Q_Z(u)du\leq0\] for any integrable $Z$ with mean zero is equivalent. After changing variables, we obtain the desired characterization. Conversely, let $X$ and $Z$ be comonotonic. For each $n\in\mathbb N$ and each $i=1,\ldots,n$, $X+\frac{i-1}{n}Z$ and $X+\frac{i}{n}Z$ are comonotonic. Calling $F_t$ the distribution of $X+tZ$, for any $t\in[0,1]$, we have 
\[
\Phi(F_1)-\Phi(F_0)=\sum_{i=1}^n\frac{1}{n}\left\{\int_0^1U_\Phi^\prime(Q_{X+\frac{i-1}{n}Z}(u),F_{\frac{i-1}{n}})Q_{Z}(u)du\right\}+o\left(\frac{1}{n}\right).
\] The terms in brackets are non positive, hence, letting $n\rightarrow\infty$, we have $\Phi(F_t)-\Phi(F_0)\leq0$, which yields the result by Proposition~\ref{proposition: LM}.
\end{proof}

We now show how this notion of Bickel-Lehmann dispersion and the Landsberger-Meilijson characterization can be extended to multivariate prospects and how it can be applied to the ranking of risk aversion of multivariate rank dependent utility maximizers. To that end, we appeal to the multivariate notions of quantiles and comonotonicity developed in \cite{GH:2011}, \cite{EGH:2009} and \cite{PS:08}.

\subsection{Local utility and multivariate mean preserving increases in risk}
\label{subsection:multiMMPIR}

\subsubsection{Multivariate quantiles and comonotonicity}
\label{subsection: multivariate quantiles and comonotonicity}

Ekeland, Galichon and Henry \cite{GH:2011}, \cite{EGH:2009} define multivariate quantiles by extending the variational characterization of univariate quantiles based on rearrangement inequalities of Hardy, Littlewood and P\'{o}lya \cite{HLP52}. The following well known equality
\begin{equation}
\int_{0}^{1}Q_{X}(u)u\;du=\max \left\{ \mathbb{E}[X\tilde U]:\;\tilde U%
\mbox{
uniformly distributed on }[0,1]\right\} ,  \label{equation: polya}
\end{equation}
is extended to the multivariate case in the following way. Let $\mu$ is a reference absolutely continuous distribution on $\mathbb R^d$ with finite second moment. This could be, for instance, the uniform distribution on the unit hypercube in $\mathbb R^d$. Let $X$ be a random vector in $\mathcal D$.  The quantile $Q_X$ of $X$ is defined as the version of $X$ (i.e., random vector with the same distriution as $X$), which maximizes correlation with a random vector $U=_d\mu$:
\begin{equation}
\mathbb{E}[Q_X(U)\cdot U]=\max\left\{ \mathbb{E}[X\cdot\tilde U]:\;\tilde{U}%
=_d\mu \right\}.
\label{equation: multivariate quantile}
\end{equation}
It follows
from the theory of optimal transportation (see Theorem 2.12(ii), p.
66 of \cite{Villani:2003}) that there exists an essentially unique
convex lower semi-continuous function $V:{\mathbb{R}}^{d}\rightarrow {%
\mathbb{R}}$ such that $Q_X=\nabla V$ satisfies Equation~\ref{equation: multivariate quantile}. Hence the definition of multivariate quantiles due to \cite{GH:2011} and \cite{EGH:2009}.
\begin{definition}[$\protect\mu $-quantile]
The $\mu $-quantile function of a random vector $X$ in $\mathcal D$ with
respect to an absolutely continuous distribution $\mu $ on $\mathbb{R}^{d}$
is defined by $Q_X$ in Equation~(\ref{equation: multivariate quantile}). \label{definition: generalized quantiles}
\end{definition}
This concept of a multivariate quantile is the counterpart of the
definition of multivariate comonotonicity in \cite{GH:2011} and \cite{EGH:2009}, motivated by the fact that two univariate prospects $X$ and $Y$ are comonotonic if there is a prospect $U$ and non-decreasing maps $T_{X}$ and $T_{Y}$ such that $Y=T_{Y}(U)$ and $%
X=T_{X}(U)$ almost surely or, equivalently, $\mathbb{E}[UX]=\max \left\{
\mathbb{E}[\tilde{U}X]:\;\tilde{U}=_{d}U\right\} $ and $\mathbb{E}[UY]=\max
\left\{ \mathbb{E}[\tilde{U}Y]:\;\tilde{U}=_{d}U\right\} $. \begin{definition}[$\protect\mu$-comonotonicity]
Random vectors $X$ and $Y$ in $\mathcal D$ are called
$\mu $-comonotonic if there exists $U=_d\mu$ such that
$\mathbb{E}[X\cdot U]=\max \left\{
\mathbb{E}[\tilde X\cdot U]:\;\tilde X=_{d}X\right\}$ and $\mathbb{E}[Y\cdot U]=\max \left\{
\mathbb{E}[\tilde Y\cdot U]:\;\tilde Y=_{d}Y\right\}$.
\label{definition: mu comonotonicity}
\end{definition}
Two random vectors are $\mu$-comonotonic if they can be rearranged simultaneously
so that they are both equal to their $\mu$-quantile. Another variational notion of multivariate comonotonicity, called $c$-comonotonicity, is proposed in Puccetti and Scarsini \cite{PS:08}.
\begin{definition}[$c$-comonotonicity]
Random vectors $X$ and $Y$ in $\mathcal D$ are called
$c$-comonotonic if there exists a convex function $V$ such that
$Y=\nabla V(X)$.\label{definition: c comonotonicity}
\end{definition}
Both $\mu$-comonotonicity and $c$-comonotonicity will feature in the extension of Bickel-Lehmann dispersion in the following section.

\subsubsection{Multivariate mean preserving increases in risk}
\label{subsection: muBL}
The Bickel-Lehmann dispersion order and its mean-preserving version in \cite{Quiggin:92}, monotone MPIR, rely on the notion of monotone single crossings, hence on the monotonicity of the function $Q_Y-Q_X$. A natural extension of the class of non-decreasing functions to functions on $\mathbb R^d$ is the class of gradients of convex functions, whose definition doesn't rely on the ordering on the real line. Hence the following definition of $\mu$-Bickel-Lehmann dispersion, which depends on the baseline distribution $\mu$ relative to which multivariate quantiles are defined.
\begin{definition}[$\mu$-Bickel-Lehmann dispersion]
A random vector $X\in\mathcal D$ is called $\mu$-Bickel-Lehmann less dispersed than a random vector $Y\in\mathcal D$, denoted $X\precsim_{\mu BL}Y$, if there exists a convex function $V:\mathbb R^d\rightarrow\mathbb R$ such that the $\mu$-quantiles $Q_X$ and $Q_Y$ of $X$ and $Y$ satisfy $Q_Y(u)-Q_X(u)=\nabla V(u)$ for $\mu$-almost all $u\in[0,1]^d$.\label{definition:mu-BL dispersion}\end{definition}
As defined above, $\mu$-Bickel-Lehmann dispersion defines a transitive binary relation, and therefore an order on $\mathcal D$. Indeed, if $X\precsim_{\mu BL}Y$ and $Y\precsim_{\mu BL}Z$, then $Q_Y(u)-Q_X(u)=\nabla V(u)$ and $Q_Z(u)-Q_Y(u)=\nabla W(u)$. Therefore, $Q_Z(u)-Q_X(u)=\nabla (V(u)+W(u))$ so that $X\precsim_{\mu BL}Z$. When $d=1$, this definition simplifies to Definition~\ref{definition: BL dispersion}.

\subsubsection{Characterization} We have the following generalization of the Landsberger-Meilijson characterization of Proposition~\ref{proposition: LM}.
\begin{theorem}\label{theorem: muBL}
$\mbox{ }$A random vector $X\in\mathcal D$ is $\mu$-Bickel-Lehmann less dispersed than a random vector $Y\in\mathcal D$ if and only if there exists a random vector $Z\in\mathcal D$ such that (i) $X$ and $Z$ are $\mu$-comonotonic and (ii) $Y=_dX+Z$.
\end{theorem}
\begin{proof}[Proof of Theorem~\ref{theorem: muBL}]
Assume $X\precsim_{\mu BL}Y$ and call $Q_X$ and $Q_Y$ the $\mu$-quantiles of $X$ and $Y$. Let $U$ be a random vector with distribution $\mu$ such that $X=Q_X(U)$. By assumption, $\nabla V(U)$ is equal to $Q_Y(U)-Q_X(U)=Q_Y(U)-X$. Call $Z=\nabla V(U)$. By Theorem 2.12(ii), p.
66 of \cite{Villani:2003}, $\nabla V$ is the $\mu$-quantile $Q_Z$ of $Z$. Hence we have $X=Q_X(U)$ and $Z=Q_Z(U)$ and $X$ and $Z$ are therefore $\mu$-comonotonic and we have $Y=_dQ_Y(U)=X+Z$ as required. Conversely, take $X$ and $Z$ $\mu$-comonotonic. Then $X=Q_X(U)$ and $Z=Q_Z(U)$ for some $U=_d\mu$, where $Q_X$ and $Q_Z$ are the $\mu$-quantiles of $X$ and $Z$ respectively. Call $Y=X+Z$ and $Q_Y=Q_{X+Z}$ the $\mu$-quantile of $Y$. In the proof of Theorem~1 of \cite{GH:2011}, it is shown that $Q_{X+Z}=Q_X+Q_Z$ when $X$ and $Z$ are $\mu$-comonotonic. Hence, we have $Q_Y=Q_X+Q_Z$, i.e., $Q_Y-Q_X=Q_Z$, and $Q_Z$ is the gradient of a convex function by Definition~\ref{definition: generalized quantiles}. The result follows.\end{proof}

The characterization given in Theorem~\ref{theorem: muBL} now allows us to generalize our characterization of MMPIR aversion to the multivariate case.

\begin{proposition}[Local utility of multivariate MMPIR averse decision makers] A decision functional $\Phi$ satisfying Assumption~\ref{ass:fre}, is $\mu$-MMPIR averse if and only if its local utility
function $U_\Phi $ satisfies \[\mathbb E_\mu\left[\nabla V(U)\cdot\nabla
U_\Phi (Q_X(U);F_X)\right]\leq0\] for all $V$ convex with
$\mathbb E_\mu V(U)=0$ and all $X\in\mathcal D$ with distribution function $F_X$ and $\mu$-quantile function $Q_X$.\label{proposition: mu-MMPIR aversion}
\end{proposition}

\begin{proof}[Proof of Proposition~\ref{proposition: mu-MMPIR aversion}]
Let $Y$ dominate $X$ with respect to mean preserving
$\mu$-Bickel-Lehmann dispersion, i.e., $Y\succsim_{\mu-MMPIR}X$.
This is equivalent to $Y=_dX+Z$ with $X$ and $Z$
$\mu$-comonotonic, $\mathbb EZ=0$. For each $\varepsilon>0$, define $Y_\varepsilon=X+\varepsilon Z$, which also dominates $X$ with respect to $\mu$-Bickel-Lehmann dispersion. $\Phi$ is $\mu$-MMPIR
averse if and only if for all $\varepsilon>0$, $\Phi(X+\varepsilon Z)-\Phi(X)\leq0$. Denoting $Q_{X+\varepsilon Z}$ and $Q_X$ the
$\mu$-quantiles of $Y_\varepsilon$ and $X$ respectively and $U=_d\mu$, comonotonicity of $X$ and $Z$ implies $Q_{Y_\varepsilon}(U)=Q_{X+\varepsilon Z}(U)=Q_X(U)+\varepsilon Q_Z(U).$ 

Now, calling $F_\varepsilon$ the distribution function of $Y_\varepsilon$ and $F$ the distribution function of $X$, we have by Assumption~\ref{ass:fre},
\begin{eqnarray*}\Phi(F_\varepsilon)-\Phi(F)&=&\int U_\Phi (x,F)[dF_\varepsilon(x)-dF(x)]+o(\varepsilon)\\
&=&\int [U_\Phi (Q_{X+\varepsilon Z}(u))-U_\Phi (Q_X(u))]d\mu(u)+o(\varepsilon)\\
&=&\int [U_\Phi (Q_{X}+Q_{\varepsilon Z}(u))-U_\Phi (Q_X(u))]d\mu(u)+o(\varepsilon)\\
&=&\varepsilon\int Q_Z(U)\cdot\nabla U_\Phi (Q_X(u),F)d\mu(u)+o(\varepsilon).
\end{eqnarray*}  
Therefore we have $\mathbb E[ \nabla V(U) \cdot \nabla U_\Phi (Q_X(U),F)] \leq0$ as required. The converse follows with the same reasoning as in the proof of Theorem~\ref{theorem: MMPIR aversion}.
\end{proof}

The characterization given in Theorem~\ref{theorem: muBL} is also crucial to the results in the next section on comparative risk attitudes of multivariate rank dependent utility maximizers.

\subsubsection{Relation to other multivariate dispersion orders}
We now look at the relation between $\mu$-Bickel-Lehmann dispersion and other generalizations of Bickel-Lehmann dispersion proposed in the statistical literature. The notion of {\em strong dispersion} was proposed by \cite{GW:95}.
\begin{definition}[Strong dispersive order] $Y$ is
said to dominate $X$ in the strong dispersive order, denoted
$Y\succsim_{SD}X$ if $Y=_d\phi(X)$, where $\phi$ is an
expansion, i.e., such that
$\|\phi(x)-\phi(x^\prime)\|\geq\|x-x^\prime\|$ for all pairs $(x,x^\prime)$.\end{definition}
The following Proposition gives conditions under which $\mu$-Bickel-Lehmann implies \cite{GW:95}'s strong dispersion.
\begin{proposition} $\mbox{ }$Let $X$ and $Y$ be two random vectors in $\mathcal D$. The following propositions hold.\begin{itemize}\item[1.] $Y$ is more dispersed than $X$ in the strong dispersion order, i.e., $Y\succsim_{SD}X$, if $Y=_dX+Z$, where $X$ and $Z$ are $c$-comonotonic. \item[2.] If $Y\succsim_{\mu BL}X$ and the $\mu$-quantiles of $X$ and $Y$ are gradients of strictly convex functions, then $Y\succsim_{SD}X$.\end{itemize} \label{proposition: strong dispersion}\end{proposition}
\begin{proof}[Proof of Proposition~\ref{proposition: strong dispersion}] If $Y\succsim_{\mu-BL}X$, then by Theorem~\ref{theorem: muBL},
$Y=_dX+Z$, where $X$ and $Z$ are $\mu$-comonotonic. Hence
\[Y=_dQ_{X+Z}(U)=Q_X(U)+Q_Z(U)=_dX+Q_Z(Q_X^{-1}(X)),\] where
$Q_X=\nabla V_X$ and $Q_Z=\nabla V_Z$ are gradients of convex
functions. Therefore, denoting $\phi(x)=x+\psi(x)=x+\nabla
V_Z\circ(\nabla V_X)^{-1}(x)$, we need to show that $\phi$ satisfies
the condition $J_\phi^T(x)J_\phi(x)-I\geq0$ for all $x$ as in the
characterization of the strong dispersive order in Theorem~2 of \cite{GW:95}. This
follows from the fact that the jacobian of a gradient of a strictly convex
function is symmetric positive definite. Now, if two matrices $S_1$ and $S_2$ are both symmetric and positive definite, then, so is $S_1^{1/2}S_2S_1^{1/2}$. The latter is therefore diagonalizable with positive eigenvalues. Since $S_1^{1/2}S_2S_1^{1/2}x=\lambda x$ is equivalent to $S_1S_2 y=\lambda y$ with $y=S_1^{1/2}x$,  $S_1S_2$ has the same eigenvalues as $S_1^{1/2}S_2S_1^{1/2}$. Hence
\[J_\psi(x)=\left[J_{\nabla V_X}\left((\nabla V_X)^{-1}(x)\right)\right]^{-1}
\left[J_{\nabla V_Z}\left((\nabla V_X)^{-1}(x)\right)\right]\] has positive eigenvalues (see also Lemma~6.2.8 page 144 of \cite{AGS:2000}). This completes the proof of (ii). The proof of (i) follows the same lines with $Y=_dX+Q_Z(X)$, where $Q_Z$ is the gradient of a convex function.
\end{proof}

\subsubsection{Partial insurance and monotone mean preserving decreases in risk}
The characterization of monotone mean preserving increase in risk given in Theorem~\ref{theorem: muBL} allows us to extend to multivariate risk sharing a celebrated result of Landsberger and Meilijson in \cite{LM:94} stating that partial insurance contracts are Pareto efficient relative to second order stochastic dominance if and only if they involve a decrease in Bickel-Lehmann dispersion. Consider an individual $A$ bearing a risk $Y$ that she considers sharing with individual
$B$, in the sense that $A$ would bear $X_A$ and $B$ would bear $X_B$ with $X_A+X_B=Y$. The partial insurance contract is therefore a (potential) decrease in the risk borne by $A$ from $Y$ to $X_A$.
The new allocation $(X_A,X_B)$ is shown in \cite{CDG:09} (up to technical regularity conditions) to be Pareto efficient (in the sense that it can't be improved for both parties irrespective of their mean preserving increase in risk averse preferences) if and only if it is $\mu$-comonotonic in the sense of Definition~\ref{definition: mu comonotonicity}. Now, by Theorem~\ref{theorem: muBL}, $\mu$-comonotonicity of $X_A$ and $X_B$, with $X_A+X_B=Y$, is equivalent to $X_A$ being Bickel-Lehmann less dispersed than $Y$. We therefore recover the strong relation between Quiggin's notion of monotone mean preserving increases in risk and partial insurance identified in \cite{LM:94} and extend it to multivariate risk sharing.

\section{Increasing risk aversion and rank dependent utility}
\label{subsection: IRA}
To make interpersonal comparisons of attitudes to multivariate risk, we define compensated increases in risk in the spirit of \cite{DS:74}.
\begin{definition}[Compensated Increases in Risk]
Let $\Phi$ be the functional representing a decision maker's preferences over multivariate prospects in $\mathcal D$. A prospect $Y\in\mathcal D$ is a compensated increase in risk from the point of view of $\Phi$ if $X\precsim_{\mu BL}Y$ and $\Phi(Y)=\Phi(X)$.\end{definition}
A ranking of risk aversion is then derived in the usual way, except that the ranking of aversion to multivariate risks is predicated on the reference measure $\mu$ in the definition of dispersion.
\begin{definition}[Increasing risk aversion]
A decision maker $\tilde\Phi$ is more risk averse than a decision maker $\Phi$ if $\tilde\Phi$ is averse to a compensated increase in risk from the point of view of $\Phi$, i.e., if $X\precsim_{\mu BL}Y$ and $\Phi(Y)=\Phi(X)$ imply $\tilde\Phi(Y)\leq\tilde\Phi(X)$.
\end{definition}
In the special case of rank dependent utility maximizers, aversion to monotone MPIR and increasing risk aversion take a very simple form. We consider here the multivariate generalization of Yaari decision makers given in \cite{GH:2011}. A multivariate rank dependent utility maximizer is characterized by a functional $\Phi$ on multivariate prospects $X\in\mathcal D$, which is a weighted sum of $\mu$-quantiles, i.e., \begin{eqnarray}\label{equation: MultiYaari} \Phi(X)=\mathbb E[Q_X(U)\cdot\phi(U)],\end{eqnarray}
where $Q_X$ is the $\mu$-quantile of $X$, $U=_d\mu$ and $\phi(U)\in\mathcal D$.
As shown in Theorem~1 of \cite{GH:2011}, $\Phi(X+Z)=\Phi(X)+\Phi(Z)$ when $X$ and $Z$ are $\mu$-comonotonic. Hence we immediately find the following characterization of monotone MPIR aversion and increasing risk aversion.
\begin{theorem}[Rank dependent utility]\label{theorem: RDU}
Let $\Phi$ and $\tilde\Phi$ be multivariate rank dependent utility functionals, i.e., $\Phi$ and $\tilde\Phi$ satisfy (\ref{equation: MultiYaari}). Then the following hold.
\begin{itemize}\item[(a)] $\Phi$ is averse to a monotone MPIR (i.e., a mean preserving $\mu$-Bickel-Lehmann dispersion) if and only if for all $Z\in\mathcal D$, $\Phi(Z)\leq\Phi(\mathbb EZ)$. \item[(b)] $\tilde\Phi$ is more risk averse than $\Phi$ iff for all $Z\in\mathcal D$, $\Phi(Z)=0\Rightarrow\tilde\Phi(Z)\leq0$.\end{itemize}
\end{theorem}
It turns out, therefore, that aversion to MMPIR in the multivariate rank dependent utility model is equivalent to weak risk aversion ($\mathbb EX$ preferred to $X$). Since Theorem~2 of \cite{GH:2011} shows that aversion to MPIR in the multivariate RDU model is equivalent to $\phi(u)=-\alpha u+u_0$, with $\alpha>0$ and $u_0\in\mathbb R^d$, we recover the fact that MPIR averters are also monotone MPIR averters as in the univariate case.
\begin{corollary} $\mbox{ }$If $\Phi$ is averse to mean preserving increases in risk, than it is also averse to monotone mean preserving increases in risk. \end{corollary}
Yaari's rank dependent utility maximizers over stochastically independent multivariate risks in \cite{Yaari:86} are special cases of (\ref{equation: MultiYaari}) where the reference distribution $\mu$ has independent marginals. In that special case, (a) of Theorem~\ref{theorem: RDU} is equivalent to concavity of the local utility function in (\ref{equation: Yaari LU}) (i.e., non-increasing $\phi_i$ for each $i$) and (b) of Theorem~\ref{theorem: RDU} is equivalent to $\tilde\phi_i$ being a decreasing transformation of $\phi_i$ for each $i$, so that we recover the classical results of \cite{Yaari:86}.

\section*{Conclusion}
Attitudes to multivariate risks were characterized using Machina's local utility in a framework, where objects of choice are multidimensional prospects. Aversion to mean preserving increases in multivariate risk is characterized by concavity of the local utility function as in the univariate case. Comparative attitudes are characterized within the multivariate extension in \cite{GH:2011} of rank dependent utility with the help of a multivariate extension of Quiggin's monotone mean preserving increase in risk notion and a generalization of its characterization in \cite{LM:94}. This allows us to extend Landsberger and Meilijson's result on the efficiency of partial insurance contracts for monotone mean preserving reductions in risk in \cite{LM:94}. Characterization and derivation of risk premia within the multivariate rank dependent utility model is the natural next step in this research agenda.

\newpage

\bibliographystyle{ormsv080}

\end{document}